\documentclass[12pt, a4paper]{elsarticle}

\usepackage{amssymb}

\usepackage{lineno}
\usepackage{soul}
\usepackage{float}
\usepackage[hyphens]{url}
\restylefloat{table}
\usepackage{listings}
\usepackage{xcolor}
\usepackage{amsmath}
\usepackage{tabularx}
\usepackage{siunitx}
\usepackage{caption}
\usepackage{rotating}

\usepackage{geometry}
\usepackage{subcaption}
\usepackage{array}
\usepackage[htt]{hyphenat}

\graphicspath{{Images/}}

\definecolor{codegreen}{rgb}{0,0.6,0}
\definecolor{codegray}{rgb}{0.5,0.5,0.5}
\definecolor{codepurple}{rgb}{0.58,0,0.82}
\definecolor{backcolour}{rgb}{0.95,0.95,0.92}

\newcolumntype{L}{>{\arraybackslash}p{0.7\textwidth}}

\lstdefinestyle{mystyle}{
    backgroundcolor=\color{backcolour},   
    commentstyle=\color{codegreen},
    keywordstyle=\color{magenta},
    numberstyle=\tiny\color{codegray},
    stringstyle=\color{codepurple},
    basicstyle=\ttfamily\footnotesize,
    breakatwhitespace=false,         
    breaklines=true,                 
    captionpos=b,                    
    keepspaces=true,                 
    numbers=left,                    
    numbersep=5pt,                  
    showspaces=false,                
    showstringspaces=false,
    showtabs=false,                  
    tabsize=2
}

\journal{}

\begin{document}


\begin{frontmatter}



\title{An Intrusion Response System utilizing Deep Q-Networks and System Partitions}


\author[utv]{Valeria Cardellini}
\author[sap]{Emiliano Casalicchio}
\author[rm3]{Stefano Iannucci}
\author[sap]{\\ Matteo Lucantonio}
\author[msu]{Sudip Mittal}
\author[msu]{Damodar Panigrahi}
\author[utv]{Andrea Silvi}

\address[utv]{University of Rome Tor Vergata}
\address[sap]{Sapienza University of Rome}
\address[rm3]{Roma Tre  University}
\address[msu]{Mississippi State University}

\begin{abstract}
Intrusion Response is a relatively new field of research. Recent approaches for the creation of Intrusion Response Systems (IRSs) use Reinforcement Learning (RL) as a primary technique for the optimal or near-optimal selection of the proper countermeasure to take in order to stop or mitigate an ongoing attack. However, most of them do not consider the fact that systems can change over time or, in other words, that systems exhibit a non-stationary behavior. Furthermore, stateful approaches, such as those based on RL, suffer the curse of dimensionality, due to a state space growing exponentially with the size of the protected system.

In this paper, we introduce and develop an IRS software prototype, named \textit{irs-partition}. It leverages the partitioning of the protected system and Deep Q-Networks to address the curse of dimensionality by supporting a multi-agent formulation. Furthermore, it exploits transfer learning to follow the evolution of non-stationary systems.
\end{abstract}

\end{frontmatter}

\section*{Code Metadata}
\label{meta-data}


\begin{table}[H]
\centering
\begin{tabular}{|l|p{6.5cm}|p{6.5cm}|}
\hline
\textbf{Nr.} & \textbf{Code metadata description} & 
\\
\hline
C1 & Current code version & V1(tag:irs-partition-v2) \\
\hline
C2 & Permanent link to code/repository used for this code version & \url{https://github.com/dpanigra/irs-partition} \\
\hline
C3 & Code Ocean compute capsule & N/A\\
\hline
C4 & Legal code license   & Apache License 2.0 \\
\hline
C5 & Code versioning system used & git \\
\hline
C6 & Software code languages, tools, and services used & Java, Python, Shell scripts  \\
\hline
C7 & Compilation requirements, operating environments \& dependencies & openjdk  11.0.13, maven\\
\hline
C8 & If available link to developer documentation/manual &  \url{https://github.com/dpanigra/irs-partition} \\
\hline
C9 & Support email for questions & dp1657@msstate.edu\\
\hline
\end{tabular}
\caption{Code metadata}
\label{code-metadata} 
\end{table}


\section{Motivation and significance}
\label{motivation}

\begin{figure}[ht]
    \centering
    \includegraphics[width=0.6\textwidth]{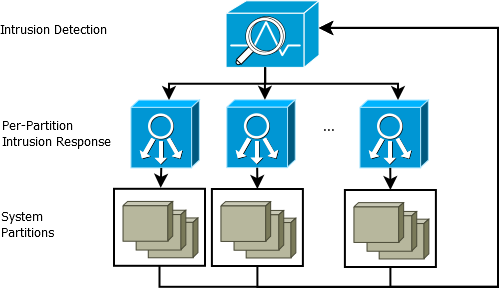}
    \caption{Role of Intrusion Detection and Intrusion Response in self-protecting systems}
    \label{fig:ids-irs}
\end{figure}

 Intrusion Detection Systems (IDSs) are widely used to detect threats to computer systems. However, they are just one of the two parts of an automatic self-protecting system, as shown in Figure~\ref{fig:ids-irs}. Indeed, while IDSs are fundamental to identify ongoing threats, they generally offer trivial response capabilities, usually based on a static mapping between the attack that has been identified and a response (e.g., Snort~\cite{snort-ips}). Unfortunately, such an approach exhibits evident limitations, such as, scalability~\cite{iannucci2016probabilistic} and lack of generalizability~\cite{iannucci2020hybrid}. For this reason, in the last decade, research on Intrusion Response Systems (IRSs) started to gain traction. The purpose of an IRS is to automatically identify the proper response to an ongoing attack, usually exploiting additional knowledge of the behavior of the attacker and of the protected system.
 
 We conducted an investigation of existing IRSs where software implementations or system design is publicly available to study how they operate on a modern computer system (e.g., ~\cite{foo2005adepts, douligeris2004ddos, koutepas2004distributed,snort-ips,ryutov2003integrated,armstrong2003autonomic, armstrong2003controller, kreidl2004feedback}). We found that all of them assume that the behavior and the topology of the protected system does not change over time or, in other words, that the protected system is \textit{stationary}. Indeed, most IRSs (e.g., \cite{foo2005adepts, armstrong2003autonomic, armstrong2003controller, kreidl2004feedback}) use either a \textit{rule-based static configuration} or a combination of static attacker and system models (e.g., \cite{tanachaiwiwat2002adaptive, toth2002evaluating}) to formulate a set of responses for the entire system. However, modern systems exhibit a non-stationary behavior, and therefore need the ability to automatically adapt to changes while dynamically predicting a near optimal response to an intrusion. Furthermore, when sophisticated techniques are employed, such as, those that make use of state information of the protected system or of the attacker, the IRS has to deal with the problem known as \textit{curse of dimensionality}, given by the exponential relationship between the size of the system (or of the attacker) model and the resulting state space.

For this reason, in this work we describe as our \textit{main contribution} an open-source licensed software prototype that implements an IRS, named \textit{irs-partition}, which builds upon the methodology introduced in~\cite{iannucci2020hybrid}.
It uses Deep Q-Learning \cite{mnih2013playing}, Reinforcement Learning (RL)~\cite{sutton2018reinforcement}, and transfer learning~\cite{olivas2009handbook} to cope with the non-stationary behavior of computer systems. To address the curse of dimensionality, its formulation supports the partitioning of the system model, therefore enabling the usage of different local modeling techniques and solvers, e.g., approaches based on Markov Decision Processes, such as, Deep Q-Learning \cite{mnih2013playing} and Dynamic Programming~\cite{sutton2018reinforcement}, or other types of optimization, such as, Linear or Integer Programming. To the best of our knowledge, our IRS software implementation is the first to be released with an Apache 2.0 license. 

The rest of the paper is organized as follows: we describe the system model and the  design of its software implementation in Section~\ref{software-description}. Then, we showcase the functionalities of the developed software with a use-case based on the open-source Google's Online Boutique application ~\cite{ob} in Section~\ref{illustrative-examples}. Finally, we discuss the impact of the software followed by conclusions and future works in Section~\ref{impact} and Section~\ref{conclusion}, respectively.

\section{System model and IRS design}
\label{software-description}
We developed and published under the Apache 2.0 license an IRS prototype, named \textit{irs-partition}. Even though the software is flexible enough to support different optimization techniques for different system partitions, at the current stage of development we introduced the support for a single solver, based on Deep Q-Learning. The latter uses a \textit{training environment} to train agents that are defined on a per-partition basis. Each agent works toward the overall system goal of keeping the system \textit{secure} by predicting the near-optimal action for its \textit{partition} using a customizable Deep Q-Learning neural network.

Software dependencies of the application include \textit{Eclipse Deeplearning4J}  (DL4J) ~\cite{dl4j}, and \textit{Reinforcement Learning for Java}  (RL4J) ~\cite{rl4j}. Both are Java implementations of deep neural network algorithms and the RL framework.

\subsection{System model}
In this section we introduce the system model and its notation. The latter is summarized in Table~\ref{tab:notation}.
\label{system-model}
\begin{table}
\centering
    \begin{tabular}{|l|L|}
        \hline
        \textbf{Symbol} & \textbf{Meaning}\\
        \hline
        $i$ & A \textit{component type} \\
        \hline
        $p_i$ & A \textit{partition} corresponding to the $i$-th component type \\
        \hline
        $i_j$ & The $j$-th \textit{component} of the $i$-th component type\\
        \hline
        $p_{i_{j}}$ & The $j$-th \textit{component} of the $i$-th type of the $i$-th partition\\
        \hline
        $S$ & The computer system model\\
        \hline
        $V$ & The set of state variables of system $S$\\
        \hline
        $v_i$ & The set of state variables of component type $i$\\
        \hline
        $v_{i_{j_T}}$ & The state of the $j$-th component of type $i$ at time $T$\\
        \hline
        $p{_{i_T}}$ & The $i$-th partition state at time $T$\\
        \hline
        $S_T$ & The state of system $S$ at time $T$\\
        \hline
        $\Sigma$ & The state space\\
        \hline
        $A$ & The set of actions available to  system $S$\\
        \hline
        $A_i$ & The set of valid actions for the $i$-th  component type\\
        \hline
        $a_i$ & A valid action ($a_i\in A_i$) for the $i$-th component type \\
        \hline
        $E(a_i)$ & The execution time for action $a_i$ \\
        \hline
        $C(a_i)$ & The cost for taking action $a_i$ \\
        \hline
        $R(\cdot)$ & The reward function \\
        \hline
        $\tau$ & The termination function \\
        \hline
        $\tau_i$ & The termination function for partition $i$ \\
        \hline
    \end{tabular}
\caption{Main notation used in this paper.}
\label{tab:notation}
\end{table}

A system contains components of different types. Each \textit{component type} can be defined at a different granularity level, as deemed necessary. Examples of component types are hardware devices, virtual appliances, software modules, web servers, application servers, database servers, network switches, load balancers, and container images. We define a \textit{component} as an instance of component type. Furthermore, we define the concept of \textit{partition} as the set of all the components of a given type $i$, i.e., $p_{i} = \cup_{j=1}^m i_j$, where $i_j$ represents component $j$ of type $i$, and $m$ is the total number of components of type $i$. The system $S$ is the set of all the \textit{partitions}, that is, $S=\{{p_1, p_2, \dots, p_n}\}$, where $n$ is the total number of partitions. In addition, given any two partitions $p_a,p_b \in S$, they do not share any component, that is, $\forall a. \forall b. a \neq b \rightarrow p_a \cap p_b = \emptyset$. In other words, partitions are disjoint.
This restriction, which has been introduced to simplify the development of the prototype, has important implications: on one hand, it eases the design, development and run-time administration of the proposed prototype. On the other hand, it could not fully capture the dynamics of a complex system, if components belonging to different partitions have some interaction. As a consequence, given the current formulation, the near-optimality of the response is guaranteed only if components belonging to different partitions do not have any interaction. This limitation will be addressed in a future release of the software prototype.

\subsection{System state}
\label{system-state}
We define a set of boolean \textit{state variables} $V=\cup_{i=1}^n v_i$, where $v_i=\{v_1, v_2, \dots, v_q\}$, where each variable $v \in v_i$ defines a specific characteristic of component type $i$ and $q$ is the total number of variables used to model the state for such component type.
For example, following the case study scenario we will describe in Section~\ref{illustrative-examples}, the variable $corrupted_i$ is applied to all the components of type $i$, and its instances represent whether or not each component of type $i$ has been compromised. The set of the variable values of all the components of a given partition $i$ at a given discrete time $T$ represents the partition state, that is, $p_{i_T} =\cup_{j=1}^m v_{i_{j_T}}$. Similarly, the system state is represented by the set of the states of its component partitions, that is, $S_{T} = \cup_{i=1}^n p_{i_T}$. Finally, $S_T \in \Sigma$, where $\Sigma$ represents the state space.

\subsection{System actions}
\label{system_actions}
We define a set of \textit{actions} which, when executed on a given component $i_j$, change the state of its corresponding partition $p_{i}$, and hence the system state. Each component type $i$ of the system has its set of valid actions, i.e., $A_i = \{a_1, a_2, \dots, a_r\}$, where $r$ is the total number of actions executable on component type $i$. Furthermore, by design, we have that $\forall j. A_i=A_{i_j}$. Hence, the set of actions available to the entire system is the union of all of the actions defined for each component type, i.e., $A = \cup_{i=1}^n A_i$. Furthermore, each action is associated with a pre-condition and a post-condition. The former, $Pre(S_{T}, a_{i_j})$, where $a_{i_j} \in A_{i_j}$, determines if action $a_{i_j}$ can be executed on component $j$ of partition $i$ when the system is in state $S_{T}$. The latter modifies the partition state, taking it from $p_{i_T}$ to $p_{i_{T+1}}$, and thus from $S_T$ to $S_{T+1}$.
\subsection{Reward and termination functions}
\label{reward_function}
For each action $a_i \in A_i$, we define its \textit{execution time}, $E(a_i)$, and \textit{cost}, $C(a_i)$, as two criteria of a \textit{reward function}. The latter returns the immediate reward obtained by a reinforcement learning agent upon its execution, and it is defined as:
\begin{equation} \label{eq1}
  R(p_{i_T}, a_{i}, p_{i_{T+1}}) =\left\{
    \begin{array}{ll}
      -2, & \mbox{ \textrm{ if $p_{i_T}=p_{i_{T+1}}$}}\\
      -w_E\frac{E(a_i)}{E_{max}} -w_C\frac{C(a_i)}{C_{max}}, & \mbox{  otherwise}. \\
    \end{array}
  \right.
\end{equation}  
where $E_{max}$ and $C_{max}$ are respectively the maximum execution time and the maximum cost; $w_E, w_C \in [0,1]$ are the corresponding optimization weights. $R(p_{i_T}, a_{i}, p_{i_{T+1}})$ returns a high penalty score of $-2$ if an action, $a_{i}$, cannot be run because the preconditions are not met. This specific formulation is a technical requirement of the Deep Q-Learning solver implementation of the DL4J library. 

Finally, the \textit{termination function} is used to identify the set of states in which the system is considered \textit{secure}. We define a per-partition termination function as $\tau_i: p_{i_T} \rightarrow \{true, false\}$, and a system-level termination function as $\tau=\bigwedge_{i=1}^n\tau_i(p_{i_T})$.

\subsection{Software design}
\label{software_design}
\begin{figure}[htbp]
    \centering
    \includegraphics[width=\textwidth]{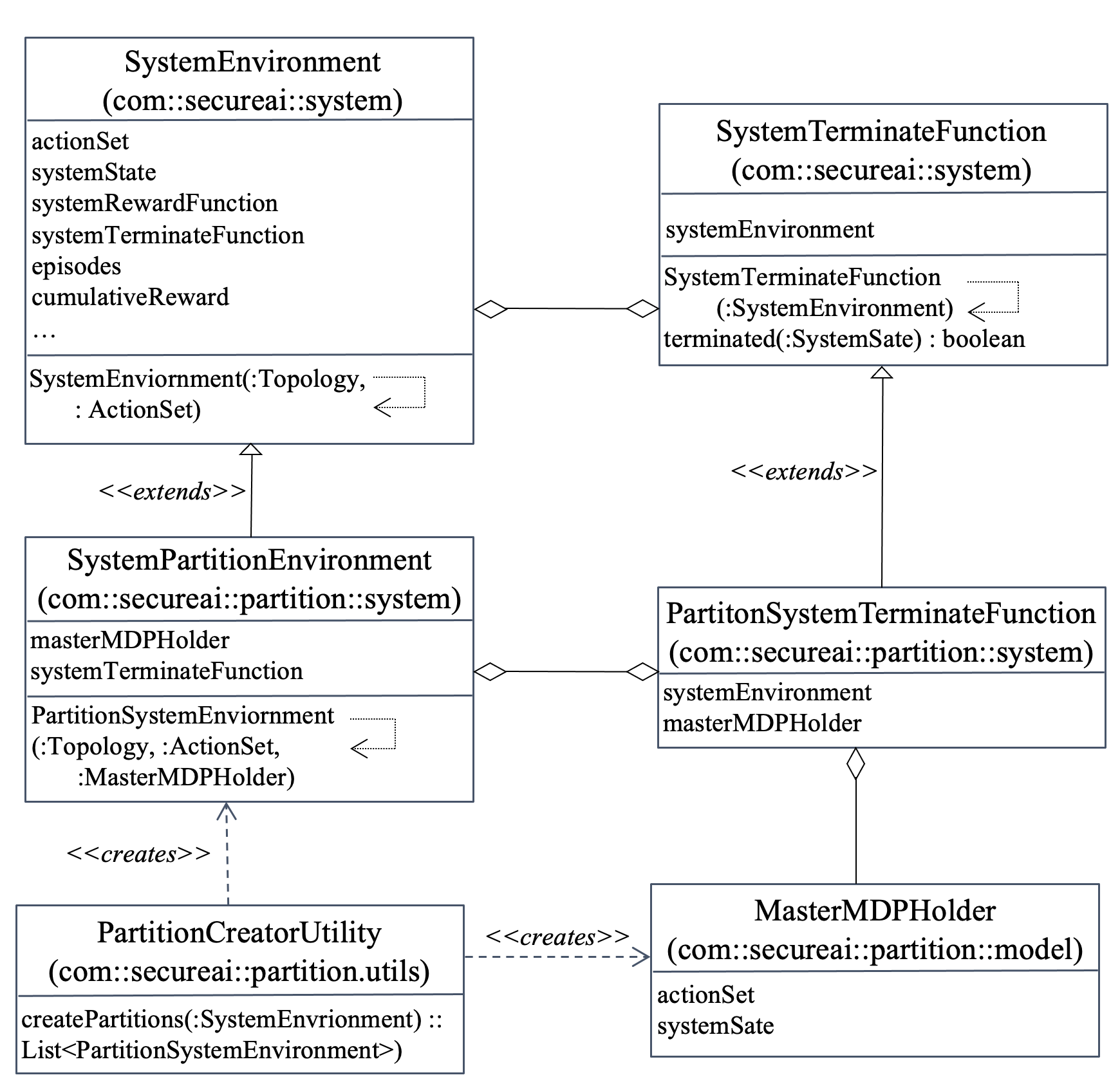}
    \caption{Class diagram of the main classes of \emph{irs-partition} software}
    \label{fig:figure1}
\end{figure}

We implement the system model $S$, system state variables $V$, actions $A$, partition state $p_{i_T}$, reward function $R$
, termination function $\tau$, and partition termination function $\tau_i$ respectively, in the \texttt{SystemEnvironment} (SE), \texttt{SystemState}, \texttt{SystemAction}, \texttt{SystemPartitionEnvironment} (PSE),  \texttt{SystemRewardFunction}, \texttt{SystemTerminateFunction}, and \texttt{PartitionSystemTerminateFunction} (PSTF) classes. We decompose the system model $S$ into multiple \textit{partitions}, where each partition stores only its own state variables and actions in \texttt{SystemPartitionEnvironment}, which is a subclass of \texttt{SystemEnvironment}.
All the partitions are then stored in the \texttt{List<PartitionSystem\-Environment>} list. We use \texttt{PartitionCreatorUtility} (PCU) to decompose the \texttt{SystemEnvironment} to multiple \texttt{PartitionSystemEnvironment} based on component type \textit{i}, as shown in Fig. \ref{fig:figure1}, which represents the class diagram of the main classes of the software.
The references to the full system state variables $V$, and action set $A$ are stored in \texttt{MasterMDPHolder}, which is a \textit{singleton object} 
 that acts as a central store and provides the state of the system at a discrete time $T$, $S_{T}$, and the set of actions, $A$, to objects of classes \texttt{SystemPartitionEnvrionment} and \texttt{PartitionSystemTerminateFunction}. 

\begin{figure}[htbp]
    \centering
    \includegraphics[width=\textwidth]{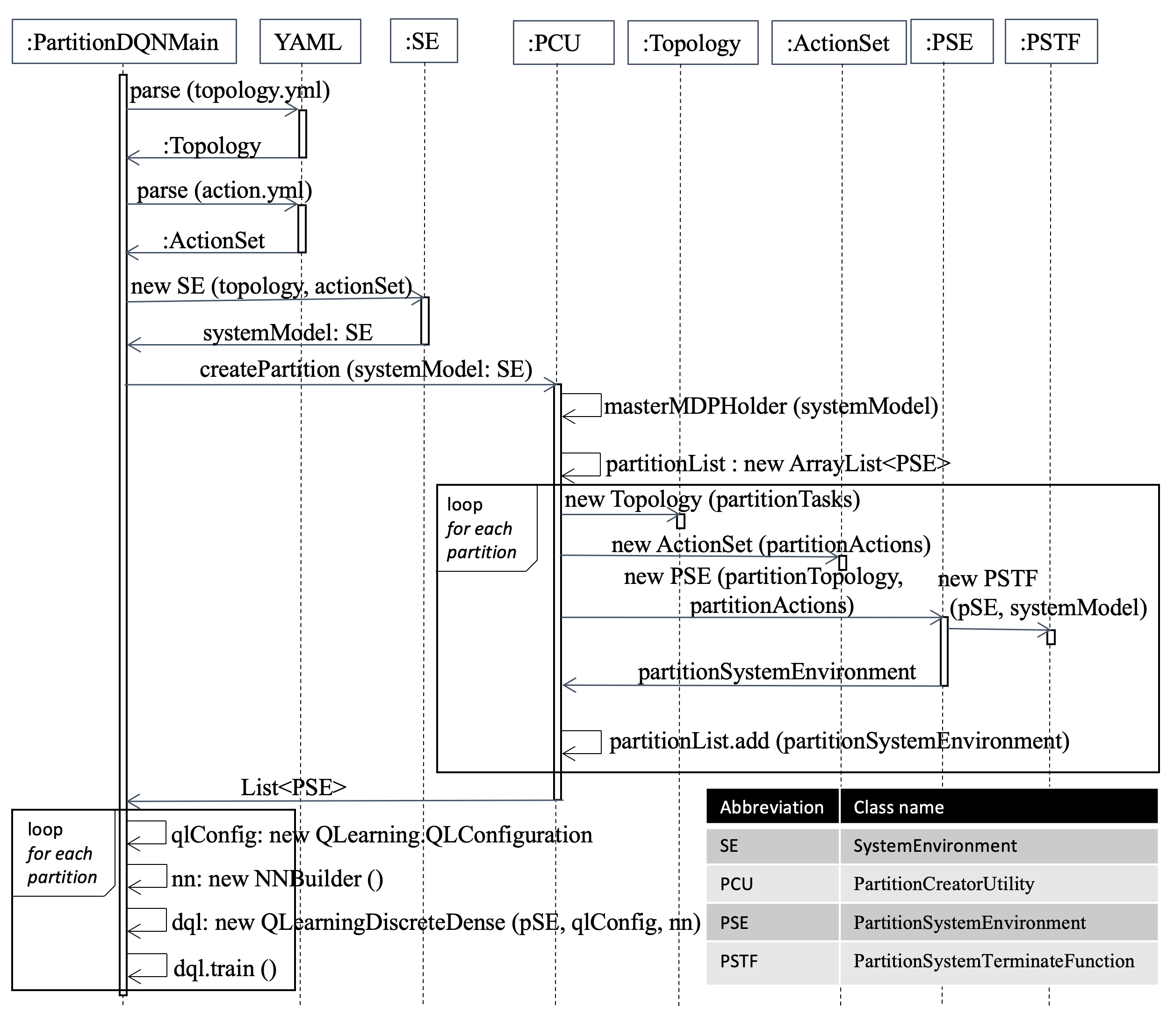}
    \caption{Sequence diagram to create deep neural nets}
    \label{fig:figure2}
\end{figure}

The execution of our software starts with the \texttt{main} function of \texttt{PartitionDQNMain}, where we create the system model ($S$) in \texttt{SystemEnvironment} from the \texttt{.yml} configuration files, store the system state ($S_T$) in \texttt{MasterMDPHolder}, decompose $S$ into partitions, store each partition in \texttt{SystemPartitionEnvironment}, and create one DNN for each partition as shown in the sequence diagram of Fig.~\ref{fig:figure2}.

We train one agent on each partition $p_i$. Each agent is responsible for providing the local near-optimal next action, 
according to the current partition state. 
Given the formulation of the system model as a set of disjoint partitions, the set of predicted optimal local actions leads to a global optimum. We use Deep Q-Learning with Monte Carlo simulation to train the agents. We utilize \texttt{QLearningDiscreteDense}~\cite{rl4j} for Deep Q-Learning with configurable parameters. The simulation begins with an initial system state configured in \texttt{SystemState} by the system administrator. Then, based on the initial state, a set of actions, \texttt{ActionSet}, (at most one for each partition) is executed on the environment, represented by \texttt{PartitionSystemEnvironment}, which 
returns a set of rewards (from \texttt{SystemRewardFunction}) and the next system state. Such actions are chosen by the agent by either exploiting the acquired knowledge, and therefore trying to maximize the expected discounted reward, or by exploring actions whose outcome, in terms of reward and transition, are still unknown. The latter case occurs with probability $\epsilon=0.01$ during the first epoch, and the parameter is gradually reduced to $0$ after $1500$ epochs. We store the state $S_T$, the action $a_{T+1}$, and the reward $R(S_T, a, S_{T+1})$ in the memory called \textit{experience}. We configured the maximum size of \textit{experience} to $5000$ in a parameter \texttt{expRepMaxSize}. Finally, the epoch continues until it either terminates when the environment reaches a \textit{secure state} (as determined by the partition termination function, \texttt{PartitionSystemTerminateFunction}) or when it reaches its maximum length (as configured in \texttt{maxStep}.) After storing a batch (configured as $128$ in \texttt{batchSize} parameter) of experiences, we train multiple DNNs, one (implemented in \texttt{NNBuilder} with parameters \texttt{layers}, \texttt{hiddenSize}, and \texttt{learningRate}) for each partition, $p_{i}$, with episodes drawn from the memory using the \textit{experience replay} technique. We run many batches of episodes to retrain the DNNs to increase accuracy in the prediction of the action.

\section{Case study: Online Boutique}
\label{illustrative-examples}
\begin{figure}[htbp]
    \centering
    \includegraphics[width=\textwidth]{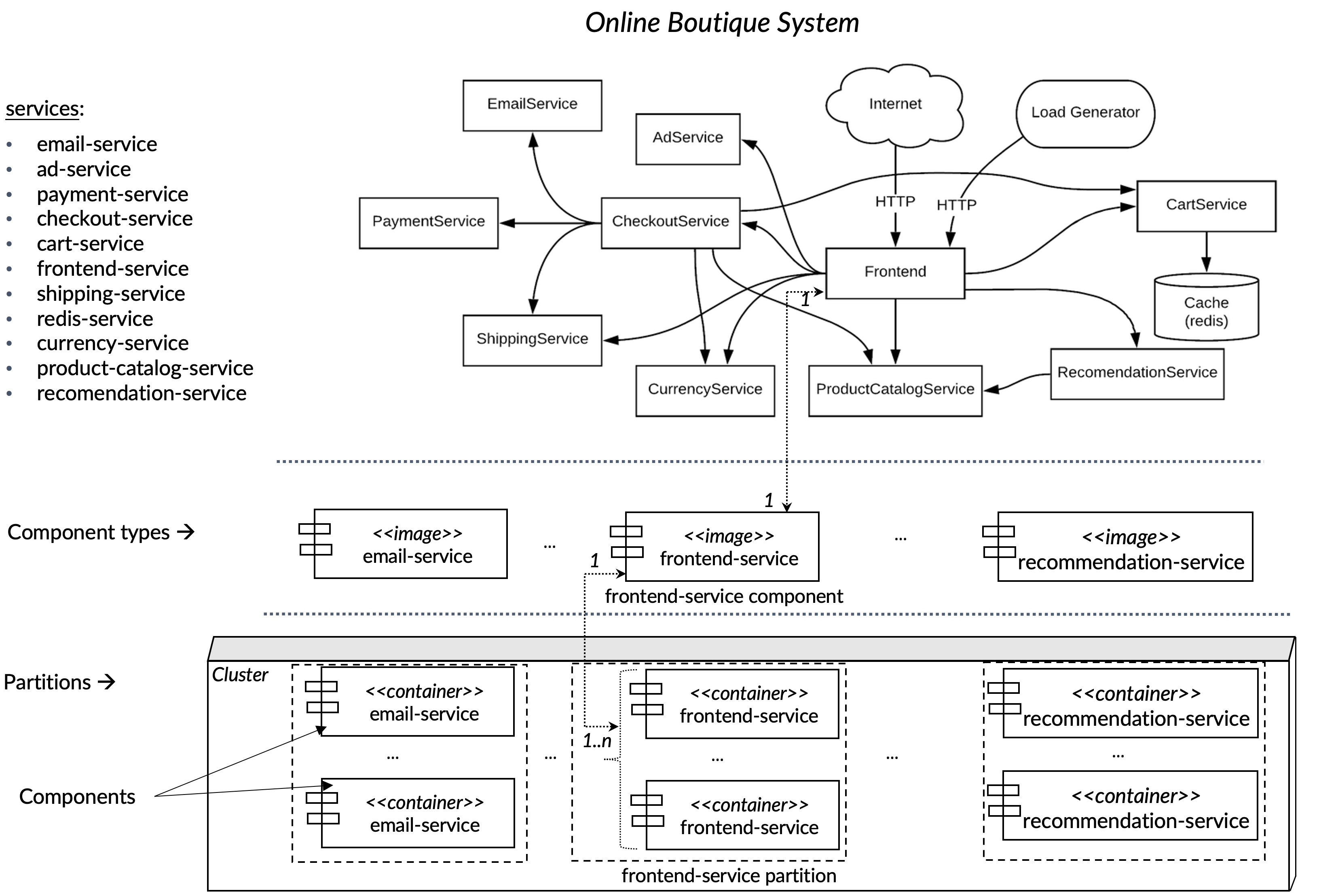}
    \caption{Architecture of the OB System}
    \label{fig:figure3}
\end{figure}

A proper validation and comparison of different IRS techniques is usually undermined by the lack of a standardized cyber-range~\cite{montemaggio2020designing}. For this reason, and in order to improve the reproducibility of our scenario and results, we illustrate the functionalities of our IRS software using a use-case scenario based on the open-source Online Boutique (OB) 2.0 system~\cite{ob}. OB is a web application used by Google to showcase cloud-enabling technologies like Kubernetes/GKE, Istio, Stackdriver, gRPC, and OpenCensus~\cite{bernstein2014containers}. It is a cloud-native application based on the microservice architectural style and is composed of 11 services, written in different languages that communicate over gRPC, plus a workload generator. It implements an online shop where users can browse items, add them to the cart, and purchase them. Fig.~\ref{fig:figure3} shows the OB system architecture, along with a representation of  a possible partitioning scheme, according to the definition of partition introduced in Section~\ref{system-model}. There are 11 partitions, one for each service. For the sake of simplicity and without loss of generality, we report experimental results showing the time needed to converge to a near-optimal solution of a scenario in which a sub-system with 2 partitions is considered. We used a machine of type~\texttt{c220g2} from CloudLab~\cite{Duplyakin+:ATC19} to run our experiments. We used the following JVM parameters: \texttt{-Xms102400m -Xmx102400m -XX:MaxMetaspaceSize=40960m}. For space reasons, we do not report experimental results on the reaction to system changes. However, the interested reader can find a detailed analysis in~\cite{iannucci2020hybrid}.

We now describe the system model of the case study and analyze 
the experiments. 
\subsection{Case study system model}
\label{experiment-system-model}
The system administrator describes the system model containing the partition information in the \texttt{topology-containers.yml} configuration file.

\begin{lstlisting} [language=Python, language=Python, caption=Configuration snippet from \textit{topology-containers.yml}, label=list:topology-listing]
frontend-service:
  replication: 1
  state:
    - start
    - active
    - restarted
    - corrupted
    - shellCorrupted
...
\end{lstlisting}

Listing~\ref{list:topology-listing} shows an example configuration of the \textit{frontend-service} partition, where the number of components in the partition is represented by the parameter \texttt{replication}, and its state variables are listed in the \texttt{state} section. This specific configuration instance shows that the component type has the following 5 state variables: \texttt{start, active, restarted, corrupted, shellCorrupted}.

For space reasons, we only list the configuration of one component type. However, we list in Table~\ref{tbl:state-variables-table} all the state variables (and their corresponding meaning) that we used to model the OB system.
 
\begin{table}
\centering
    \begin{tabular}{|l|L|}
        \hline
        \textbf{State variable} & \textbf{Meaning}\\
        \hline
        \texttt{start} & If $true$, 
        the container has started \\
        \hline
        \texttt{active} & If $true$, 
        the container is running \\
        \hline
        \texttt{corrupted} & If $true$, 
        the container is under attacker control \\
        \hline
        \texttt{restarted} & If $true$, 
        the container has been restarted after the agent requested to do so\\
        \hline
        \texttt{shellCorrupted} & If $true$, 
        the attacker has overwritten the shell /bin/sh in the container\\
        \hline
        \texttt{cartCorrupted} & If $true$, 
        the content of  Redis data store has been altered by the attacker\\
        \hline
        \texttt{confVuln} & If $true$, 
        the current configuration of Redis data store is vulnerable to potential attacks and is subject to loss of confidentiality\\
        \hline
        \texttt{intVuln} & If $true$, 
        the current configuration of Redis data store is vulnerable to potential attacks and is subject to loss of integrity\\
        \hline
        \texttt{passwordRequired} & If $true$, it mandates a password before accepting a command on Redis data store\\
        \hline
        \texttt{dangerousCmdEnabled} & If $true$, 
        dangerous commands, such as \textit{flushall}, that can potentially compromise the Redis data store, are enabled.\\
        \hline
        \texttt{accessRestricted} & If $true$, it only permits access from permitted sources, such as \textit{cart-service}, to the Redis data store.\\
        \hline
    \end{tabular}
\caption{OB System State variables list}
\label{tbl:state-variables-table} 
\end{table}

\begin{lstlisting} [language=Python, language=Python, caption=Configuration snippet from \textit{action-set-containers.yml}, label=list:action-set-listing]
start:
  execution-time: 300
  execution-cost: 100
  pre-condition: state[active] == false
  post-condition: state[active] = rand(1)
  components:
    - frontend-service
    - cart-service
    - redis-service
...
\end{lstlisting}

The administrator also defines a set of actions and provides the following parameters for each action: the reward parameters (execution time and cost), the pre-condition and the post-condition in the \texttt{action-set-containers.yml} configuration file. Listing \ref{list:action-set-listing} shows the configuration of the action \texttt{start}, consisting of: its  reward parameters (\texttt{execution-time} and \texttt{execution-cost}); the component types (\texttt{frontend-service} and \texttt{redis-service}) whose components can choose \texttt{start} as one of the action under the \texttt{components} section; the pre- and post-conditions under their respective sections. Table~\ref{tbl:actionset-table} defines all the actions along with their pre-condition, post-conditions, execution time and cost, that we modeled for the protection of the OB system.

\begin{sidewaystable*}[]
\small
\begin{tabularx}
{0.95\textwidth}
{ 
  | >{\raggedright\arraybackslash}p{0.165\textwidth} 
  | >{\raggedright\arraybackslash}p{0.2\textwidth} 
  | >{\raggedright\arraybackslash}p{0.2\textwidth} 
  | >{\raggedright\arraybackslash}p{0.2\textwidth} 
  | >{\raggedright\arraybackslash}X 
  | >{\raggedright\arraybackslash}X |}

\hline
\textbf{Action Name} & \textbf{Description} & \textbf{Pre-Condition} & \textbf{Post-Condition} & \textbf{$E(a_i)$} & \textbf{$C(a_i)$} \\ \hline \hline
$start_i$ &Start a stopped microservice & $\neg active_i$ & $P=1 \rightarrow active_i=true$ & 300 & 100 \\ \hline

$restart_i$ &Restart a malfunctioning service & $active_i \wedge corrupted_i \wedge \neg restarted_i$ & $P=0.75 \rightarrow corrupted_i=false; P=1 \rightarrow restarted_i=true$ & 500 & 300 \\ \hline

$heal_i$ &Restore a malfunctioning service from a container image & $active_i \wedge corrupted_i \vee shellCorrupted_i$ & $P=1 \rightarrow corrupted_i=false;P=1 \rightarrow shellCorrupted_i=false$ & 1000 & 500 \\ \hline

$healRedisSecure_i$ &Restore a malfunctioning Redis server from a container image & $active_i \wedge cartCorrupted_i \wedge \neg intVuln_i$& $P=1 \rightarrow cartCorrupted_i=false$ & 1000 & 500 \\ \hline

$healRedisInsecure_i$ &Restore a malfunctioning Redis server from a container image & $active_i \wedge cartCorrupted_i \wedge intVuln_i$ & $P=1 \rightarrow cartCorrupted_i=true$ & 1000 & 500 \\ \hline

$enablePassword_i$ &Configure the Redis server to request a password before a user can issue commands & $active_i\wedge \neg passwordRequired_i \wedge confVuln_i \vee intVuln_i$ & $P=1 \rightarrow passwordRequired_i=true P=1 \rightarrow confVuln_i=false; P=1 \rightarrow intVuln_i=false$ & 1000 & 1000 \\ \hline

$disableDangerousCmd_i$ &Configure the Redis server to disable dangerous commands & $active_i \wedge dangerousCmdEnabled_i\wedge intVuln_i$ & $P=1 \rightarrow dangerousCmdEnabled_i=false; P=0.85 \rightarrow intVuln_i=true$ & 50 & 500 \\ \hline

$restrictAccess_i$ &Configure firewall rules to permit access from authorized services & $active_i \wedge \neg accessRestricted_i \wedge confVuln_i \vee intVuln_i$ & $P=1 \rightarrow accessRestricted_i=true; P=0.7 \rightarrow confVuln_i=true; P=0.7 \rightarrow intVuln_i=true$ & 50 & 300 \\ \hline
\end{tabularx}
\caption{Actions list}
\label{tbl:actionset-table} 
\end{sidewaystable*}

\begin{figure}[htbp]
    \centering
    \includegraphics[width=\textwidth]{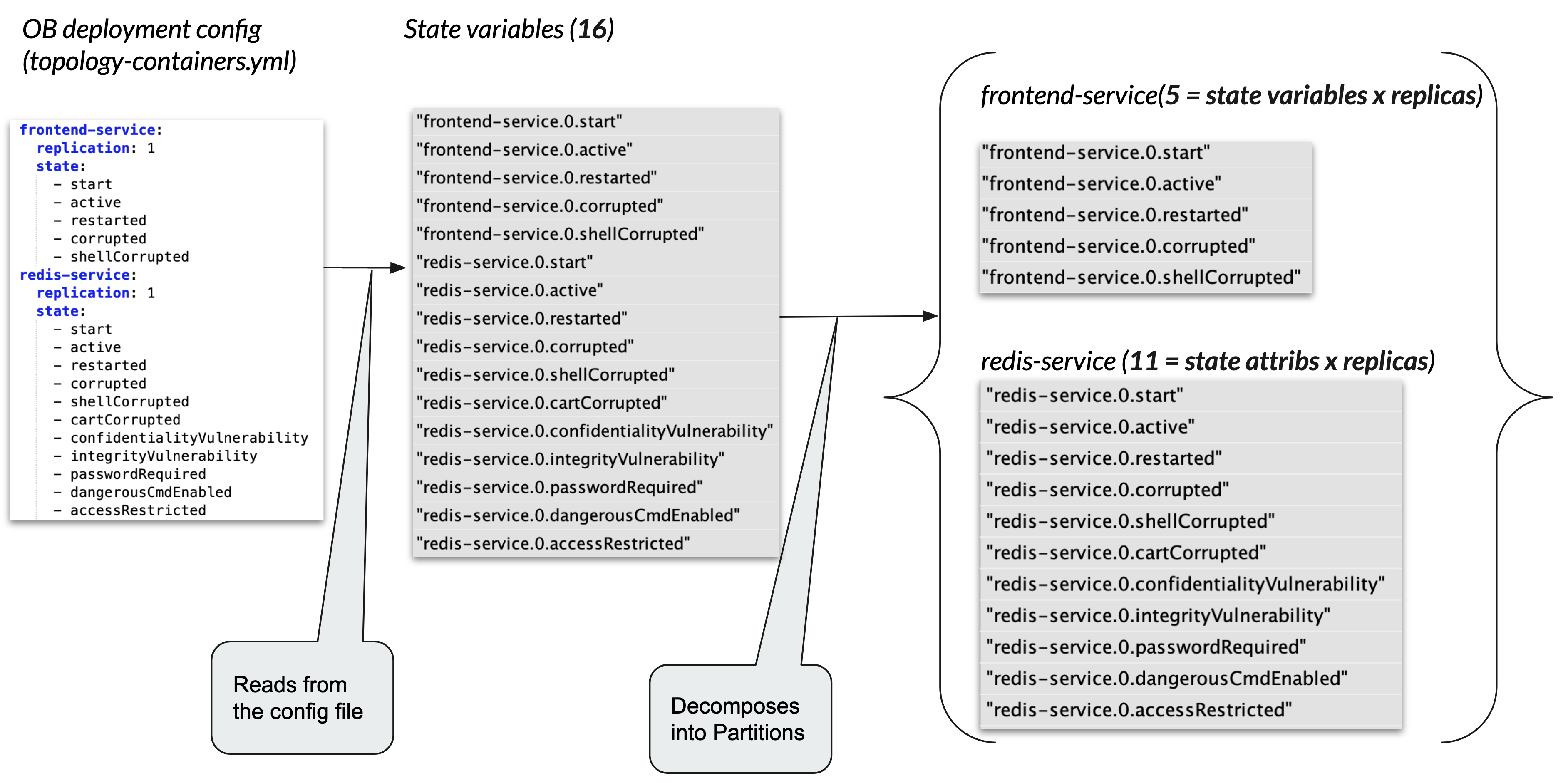}
    \caption{Relationship between the OB System state and Partition state variables}
    \label{fig:figure4}
\end{figure}

We use a total of \textit{16 state variables} and decompose the system state as shown in Fig.~\ref{fig:figure4}.
Furthermore, we implement \texttt{PartitionSystemTerminateFunction.\-terminate()} as the conjunction of the subset of the state variables reported in Table~\ref{tbl:terminal-condition}. In addition, the input to each DQN is the set of the state variable values of the its corresponding partition, and the output is one action from the set of valid actions. 
\begin{table}
\centering
    \begin{tabular}{|l|}
        \hline
        \textbf{State Variable Condition} \\
        \hline
        \texttt{active} = $true$\\
        \hline
        \texttt{corrupted} = $false$\\
        \hline
        \texttt{cartCorrupted}  = $false$\\
        \hline
        \texttt{confVuln} = $false$\\
        \hline
        \texttt{intVuln}  = $false$\\
        \hline
        \texttt{shellCorrupted} = $false$\\
        \hline
    \end{tabular}
\caption{Termination condition}
\label{tbl:terminal-condition} 
\end{table}

\begin{table}
\centering
    \begin{tabular}{|l|l|}
        \hline
        \textbf{State variable name}  & \textbf{Value} \\
        \hline
        \textit{start} & \textit{true} \\
        \hline
        \textit{active} & \textit{false} \\
        \hline
        \textit{restarted} & \textit{false} \\
        \hline
        \textit{corrupted} & \textit{true} \\
        \hline
        \textit{cartCorrupted} & \textit{true} \\
        \hline
        \textit{confVuln} & \textit{true} \\
        \hline
        \textit{intVuln} & \textit{true} \\
        \hline
        \textit{passwordRequired} & \textit{false} \\
        \hline
        \textit{dangerousCmdEnabled} & \textit{true} \\
        \hline
        \textit{accessRestricted} & \textit{false} \\
        \hline
    \end{tabular}
\caption{Initial state variable values of the OB system}
\label{tbl:initial-state-variable-values-table} 
\end{table}
\subsection{Case study experiments}
\label{experiment-analysis}
\begin{figure}[htbp]  
    \centering
    \begin{subfigure}[b]{0.45\textwidth}
        \centering
        \includegraphics[width=\textwidth]{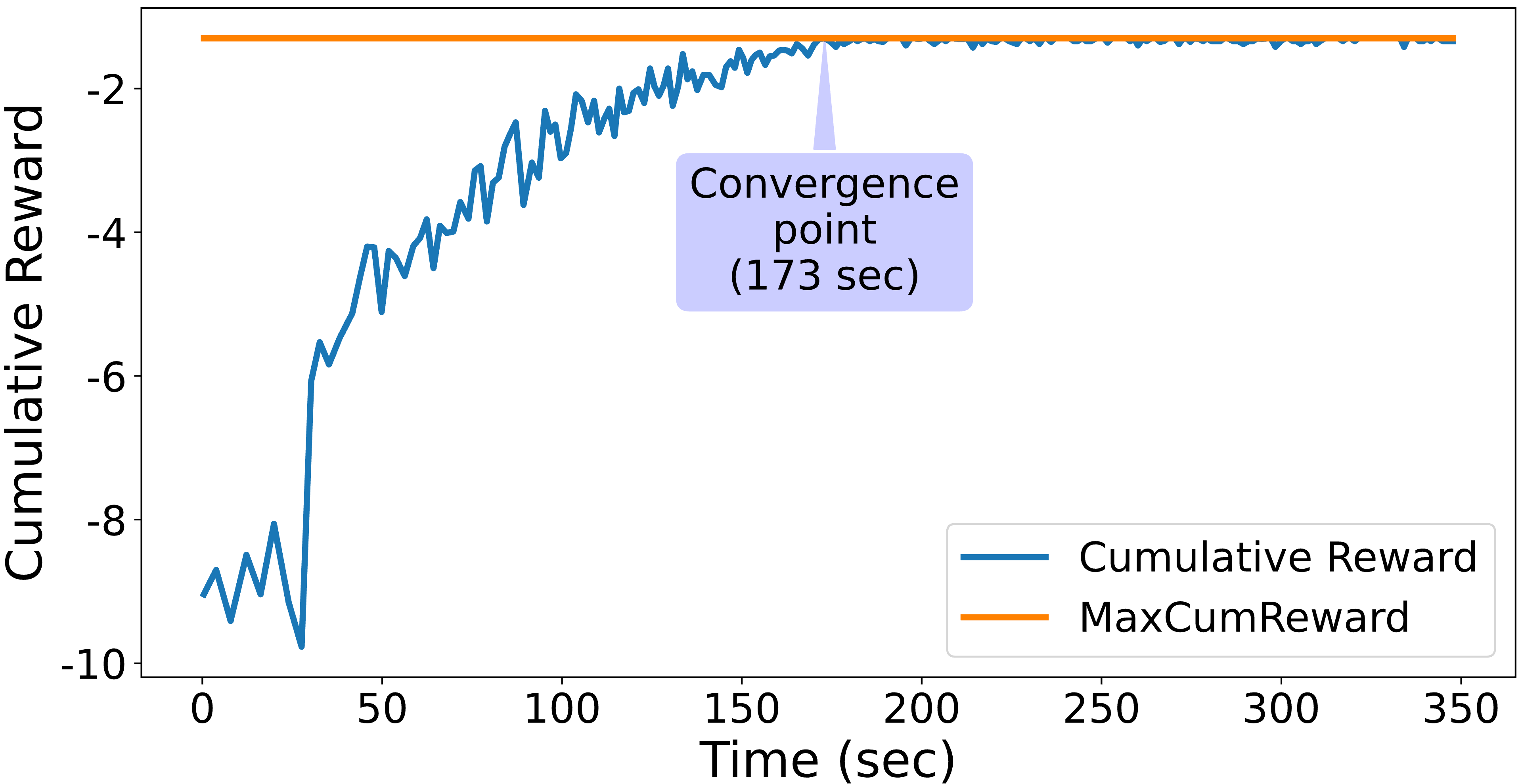}
        \caption{\textit{frontend-service} partition}\label{fig:7a}     
    \end{subfigure}
    \quad
    \begin{subfigure}[b]{0.45\textwidth}
        \centering
        \includegraphics[width=\textwidth]{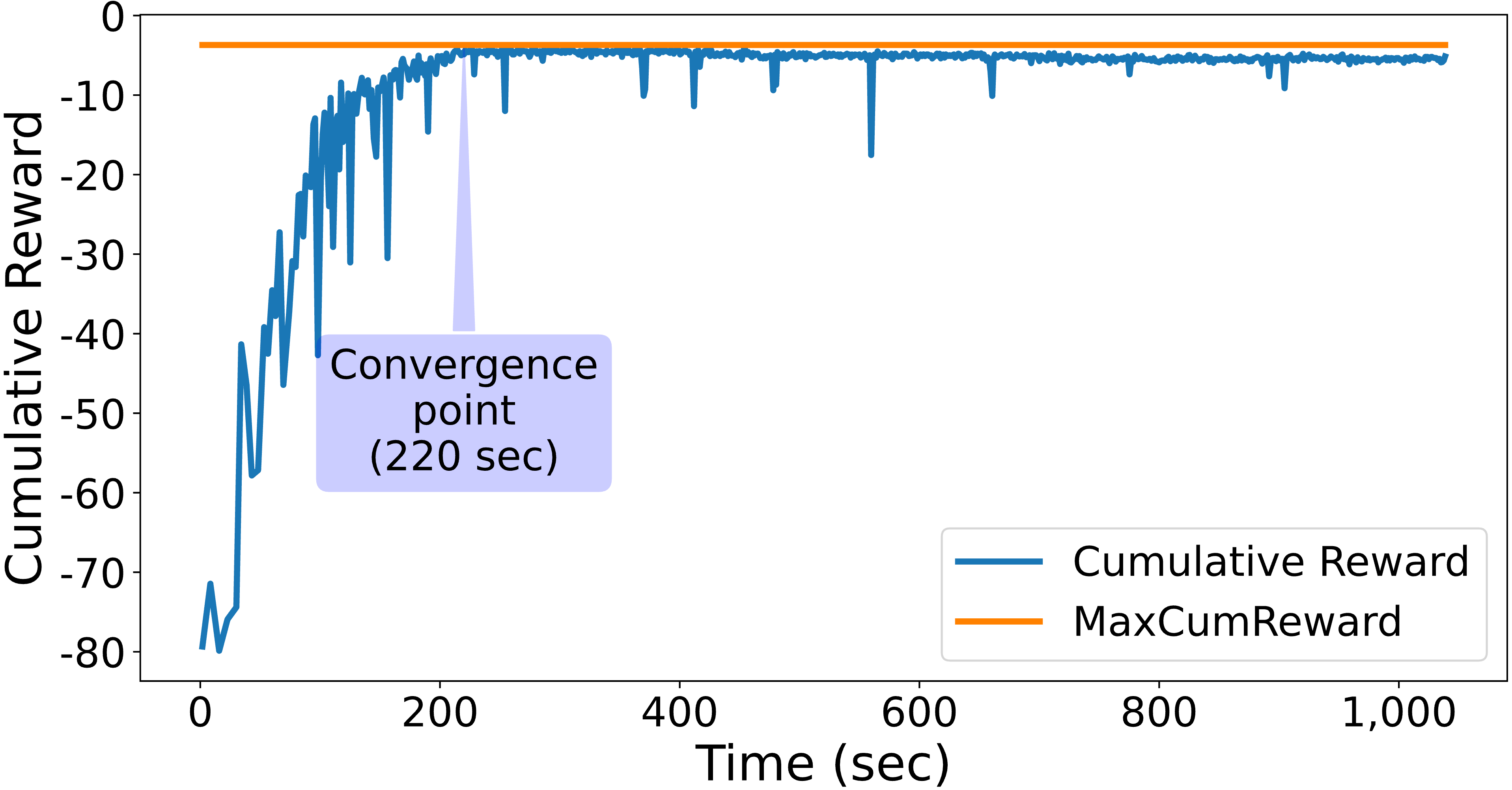}
        \caption{\textit{system} }\label{fig:7b}
    \end{subfigure}
    \caption{DQN training time vs cumulative rewards}\label{fig:figure7}
\end{figure}

We ran experiments to gather cumulative rewards in training the DQNs for the entire \textit{system} and the \textit{front-end partition}. As expected, the training time to converge to near-optimal cumulative reward of the \textit{front-end partition}, \texttt{173 sec}, is smaller than that of the \textit{system}, \texttt{220 sec}.
We calculated the optimal cumulative reward using \texttt{VIMain} and \texttt{PartitionVIMain}, which invoke the \textit{Value Iteration} implementation of RL4J. Fig.~\ref{fig:7a} and~\ref{fig:7b} respectively show the cumulative reward obtained according to the time spent on training for both,  the single \textit{front-end partition} and the \textit{system}.

\section{Impact}
\label{impact}
The \textit{irs-partition} system described in this paper further advances the state of the art in IRS software. We take a significant step forward in creating self-protecting systems that support non-stationary behavior, allow complex system partitioning, and near-optimal mitigation of local threats using multiple model types, including DQNs with customizable hyper-parameters. Our IRS software implementation with these capabilities is also the first to be released with an Apache 2.0 license. 

Our software uses a training environment with a simulated system to train the IRS agents. Thus, it makes it possible to pre-train agents in a training environment and deploy them in a live environment. We train each agent with a dedicated deep neural network,  where each network can be customized to a different architecture with its own set of hyperparameters. In addition, each agent could configure different types of modeling approaches, including DQNs, which we have used in our prototype.

\section{Conclusions}
\label{conclusion}
Cyber threats are still evolving, and the security industry needs systems that can both, detect and respond, automatically. This need requires further investigation on automatic self-protecting systems, which can help secure real-world systems exhibiting non-stationary behavior. In this paper, we demonstrated a software tool to train multiple agents in a training environment using customizable deep neural networks to build an IRS, named \textit{irs-partition}. We focused on leveraging multiple deep neural networks that predict a set of optimal actions. Moreover, the pre-trained agents immediately enhance system security using the transfer learning technique from their experience gained in a simulated system. In the future, we plan to monitor the impact and quality of the predictions, and to provide a mechanism to self-tune the deep neural networks. 

\section*{Conflict of Interest}
We declare no conflict of interests. 

\section*{Acknowledgements}
\label{acknowledgement}
All the experiments have been conducted on the NSF-sponsored CloudLab platform~\cite{Duplyakin+:ATC19}. E. Casalicchio is funded by the project Smart Defense (000090\_19RS).

\bibliographystyle{elsarticle-num}
\bibliography{irs-citations.bib}

\end{document}